\documentclass{vgtc}                          

\ifpdf
  \pdfoutput=1\relax                   
  \usepackage{graphicx}                
  \DeclareGraphicsExtensions{.pdf,.png,.jpg,.jpeg} 
\else
  \usepackage{graphicx}                
  \DeclareGraphicsExtensions{.eps}     
\fi%

\graphicspath{{figures/}{pictures/}{images/}{./}} 

\usepackage{microtype}                 
\PassOptionsToPackage{warn}{textcomp}  
\usepackage{textcomp}                  
\usepackage{mathptmx}                  
\usepackage{times}                     
\usepackage{cite}                      
\usepackage{tabu}                      
\usepackage{booktabs}                  
\usepackage[dvipsnames]{xcolor}

\onlineid{0}

\vgtccategory{Research}

\vgtcinsertpkg

\title{Exploring Head-based Mode-Switching in Virtual Reality}

\author{Rongkai Shi\thanks{e-mail: rongkai.shi19@student.xjtlu.edu.cn}, Nan Zhu\thanks{e-mail: nan.zhu18@student.xjtlu.edu.cn}, Hai-Ning Liang\thanks{\textit{Corresponding author}; e-mail: haining.liang@xjtlu.edu.cn}\\ %
    \parbox{1.7in}{\scriptsize \centering Department of Computing \\ Xi'an Jiaotong-Liverpool University}
\and Shengdong Zhao\thanks{e-mail: zhaosd@comp.nus.edu.sg}\\ %
    \parbox{1.7in}{\scriptsize \centering NUS-HCI Lab \\ National University of Singapore}}

\abstract{Mode-switching supports multilevel operations using a limited number of input methods. In Virtual Reality (VR) head-mounted displays (HMD), common approaches for mode-switching use buttons, controllers, and users' hands. However, they are inefficient and challenging to do with tasks that require both hands (e.g., when users need to use two hands during drawing operations). Using head gestures for mode-switching can be an efficient and cost-effective way, allowing for a more continuous and smooth transition between modes. In this paper, we explore the use of head gestures for mode-switching especially in scenarios when both users' hands are performing tasks. We present a first user study that evaluated eight head gestures that could be suitable for VR HMD with a dual-hand line-drawing task. Results show that move forward, move backward, roll left, and roll right led to better performance and are preferred by participants. A second study integrating these four gestures in Tilt Brush, an open-source painting VR application, is conducted to further explore the applicability of these gestures and derive insights. Results show that Tilt Brush with head gestures allowed users to change modes with ease and led to improved interaction and user experience. The paper ends with a discussion on some design recommendations for using head-based mode-switching in VR HMD.%
} 

\CCScatlist{
  \CCScatTwelve{Human-centered computing}{Human computer interaction (HCI)}{Interaction paradigms}{Virtual reality};
  \CCScatTwelve{Human-centered computing}{Human computer interaction (HCI)}{Empirical studies in HCI}{};
}

\begin{document}


\firstsection{Introduction}

\maketitle

Modes represent distinct user interface settings, where two modes lead to different results from the same input, and are manifested by how an interface responds to inputs \cite{Raskin:2000}. They are common in all types of user interfaces (UI). Mode-switching is the transition from one mode to another. However, different modes and switching between them, to an extent, can lead to errors, increased confusion, unnecessary restrictions, and added complexity in a UI \cite{Raskin:2000}. Having a practical, efficient, and easily-performed mode-switching mechanism is important for most interfaces, especially for Virtual Reality and Augmented Reality head-mounted displays (VR/AR HMD) which have more interactive possibilities and complexity.

Researchers have utilized specific hand-held devices or hand motions for mode-switching for different devices, especially for those with pen- and touch-based interfaces (e.g., \cite{Li:2005, Hinckley:2006, Tu:2012,Surale:2017}). However, mode-switching in immersive VR/AR HMD is relatively underexplored. Applications in VR HMD, just like those in other platforms, can have multiple modes. For example, a 3D-painting application (e.g., \cite{Park:2019}) can contain several modes for interacting with it, such as painting and erasing, while in a 3D-modeling application (e.g., \cite{Yu:2021}) interaction could involve object selection and manipulation, each of which has a distinct effect from the same input. Having an efficient mode-switching technique or mechanism can be supportive for scenarios that involve more than one mode. The existing research tended to focus on and suggested using hand-based interactions for mode switching \cite{Surale:2019,Smith:2019}. However, hand-based mode-switching techniques are not suitable or ideal for many VR applications, for example, when users need to use both hands for interaction, which is common in HMD \cite{LaViola:2017}. Unlike 2DUI, which have more constraints because of their input devices (e.g., mouse or pen), VR/AR provide more ways of interaction and degrees of freedom where users' hands are frequently used \cite{LaViola:2017}. As such, a hands-free approach for mode switching can be useful in many VR applications to support a richer and more continuous flow of interaction.

Voice control is one hands-free approach commonly supported by HMD. However, it has poor performance in noisy environments \cite{Pearce:2000, LaViola:2017} and is unsuitable in a public place or even a private space that is shared with others. Foot interaction is another hands-free mechanism and has been studied as an input modality in VR (e.g., \cite{Minakata:2019, Xu:2019DMove}). It can be intuitive and helpful but it also requires extra detection devices that tend to be difficult to set up \cite{Felberbaum:2018, Minakata:2019, Willich:2020}. Similarly, eye-based interaction requires additional trackers inside the HMD, can only provide limited possibilities, is sensitive to the accuracy of the eye tracker, and can lead to eye fatigue and errors \cite{Minakata:2019, Lu:2020, Yu:2021}. 

Head-based gestures can be a natural and cost-efficient hands-free approach that suits ``hands-busy'' scenarios such as those often found in painting, modeling, or gaming applications. In general, using head gestures in HMD has following benefits: (1) users are familiar with the head gestures, such as to nod, raise, or turn the head; (2) head gestures are easy, cheap, and fast to perform; (3) head movements can be tracked with high accuracy by most of the current HMD devices without any additional hardware \cite{Yan:2018, Yu:2019}. As such, head gestures might be able to aid mode-switching in a fast, low-cost, and error free manner. To best of our knowledge, there has not been prior research that has formally and systematically explored the use of head gestures for mode-switching in VR HMD. In this research, we aim to investigate what potential head gesture(s) can lead to better performance and usability for mode-switching in VR (Study One) and how they perform in a real application (Study Two).

\begin{figure*}[tb]
 \centering 
 \includegraphics[width=\linewidth]{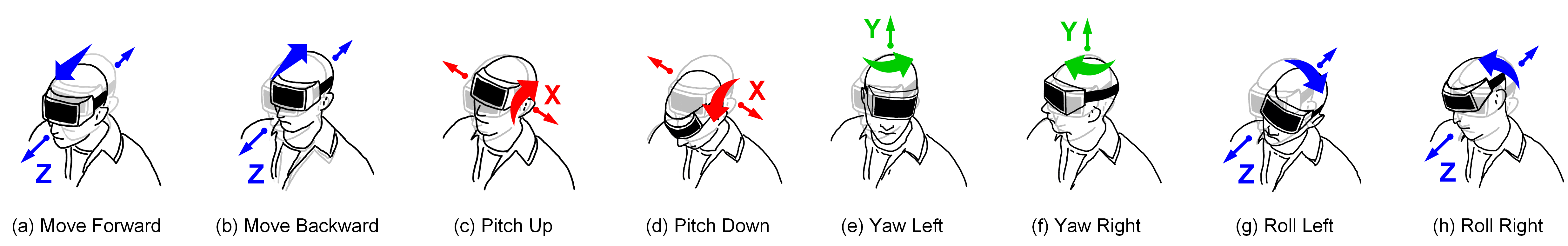}
 \caption{Investigated head-based gestures: (a) move forward, (b) move backward, (c) pitch up, (d) pitch down, (e) yaw left, (f) yaw right, (g) roll left, (h) roll right.}
 \label{fig:HeadGestures}
\end{figure*}

We first evaluated eight head gestures that were selected based on three design criteria to ensure they are potentially suitable for mode-switching in VR: move forward, move backward, pitch up, pitch down, yaw left, yaw right, roll left, and roll right (see \autoref{fig:HeadGestures}). We conducted two user studies to evaluate their usability and performance. In the first study, we collected and compared the performance and users' subjective feedback of head gestures in an extended line-drawing task which was designed based on the certified ``\textit{subtraction method}'' \cite{Dillon:1990, Li:2005, Tu:2012, Surale:2017, Surale:2019}. Results from this study show that move forward, move backward, roll left, and roll right had better performance and were preferred by the participants. In the second study, we further evaluated these four gestures in a actual VR painting application where these gestures helped achieve quick and smooth switch between the painting tools, allowing an enhanced continuous interaction and user experience with the application. 

Overall, this paper makes three main contributions: (1) it provides a first systematic exploration of using head gestures for mode-switching in VR HMD;
(2) it reports an effective integration of four head gestures in an actual application to show their applicability in allowing fast, smooth, and continuous mode-switching; 
(3) it offers a set of design considerations for incorporating head gestures for mode-switching in VR HMD. 

\section{Related Work}

\subsection{Mode-Switching in 2D User Interfaces}
The early research investigated mode switching in desktop computers. Dillon et al. \cite{Dillon:1990} measured the time cost and errors for command selection. They laid the foundation of the ``\textit{subtraction method}'' which can derive the true time spent for switching commands. Sellen et al. \cite{Sellen:1992} discussed and suggested user-maintained mechanism to reduce mode-switching errors during text editing. Kabbash et al. \cite{Kabbash:1994} evaluated two-handed input in a compound line-drawing and color selection task. Based on these prior works, Li et al. \cite{Li:2005} compared five techniques for mode switching in pen-based interfaces. They measured the mode-switching time by subtracting the time to complete single-color pie-crossing tasks from those when alternately switching between two colors. By following this approach, Hinckley et al. \cite{Hinckley:2006} investigated the performance of mode switching using a user-maintained mechanism for tablet devices. In addition, Tu et al. \cite{Tu:2012} and Surale et al. \cite{Surale:2017} followed the subtraction method and used a modified pie-crossing task to compare mode-switching techniques for pen- and touch-based interfaces, respectively. The subtraction method provides a precise measure to quantify the time cost of mode-switching and has been adapted and examined by above-mentioned work. Thus, we followed this proven, robust approach to design and conduct the experiments in our study.  

\subsection{Mode-Switching and Related Studies in HMD}
VR or AR HMD provide larger interaction space compared to 2DUI. With the support of 3D input technologies, HMD leverage the use of controller, hand gestures, head gestures, body gestures, and eye gaze \cite{LaViola:2017}. Specifically, hand gestures to do mode-switching have been discussed in HMD. Surale et al. \cite{Surale:2019} presented two studies to evaluate barehand mid-air mode-switching techniques for VR HMD. They followed the subtraction method and compared dominant and non-dominant hand gestures in a VR line-drawing task. As for AR HMD, Smith et al. \cite{Smith:2019} performed an evaluation of five mode-switching techniques and found that the non-preferred hand was fast and led to lower errors. They further evaluated the scalability of non-preferred hand mode-switching in their recent study \cite{Smith:2020}. 

The above prior work has primarily focused on exploring mode-switching techniques in controlled experimental settings. However, there has been limited exploration of such techniques in actual HMD applications. Some studies dealing with tasks in VR applications, such as menu selection \cite{Park:2019, Pfeuffer:2020}, task-switching \cite{Ens:2014,Lindlbauer:2019}, disambiguation \cite{Chen:2020}, though have involved the concept of mode-switching as part of their research, did not conduct controlled evaluations. In this research, we first conducted a systematic evaluation of mode-switching using head motions in a more controlled setting. We then incorporated the most suitable and best rated motions to a real VR application to assess their performance and usability in more uncontrolled and realistic scenario.

\subsection{Head-based Interaction}
As part of our body language, head movements are natural and deliberate, and due to this, prior work has leveraged head-based interactions in HMD. Head-gaze interaction is one of the most common approaches, which has been widely studied for target selection (e.g., \cite{Qian:2017, Blattgerste:2018, Kyto:2018, Minakata:2019}) and text entry (e.g., \cite{Yu:2017, Xu:2019proc, Xu:2019art}) in VR/AR HMD. As head movements can be performed accurately yet effectively, head-gaze interaction has been used as a standalone approach and combined with or supported by other modalities, such as eye-based interaction \cite{Qian:2017, Blattgerste:2018, Kyto:2018, Xu:2019proc, Sidenmark:2019, Sidenmark:2021, Lu:2021}. Head-gaze has also been used, for example, to interact in games \cite{Atienza:2016}, map interfaces \cite{Giannopoulos:2017}, and to construct new interaction methods \cite{Esteves:2017, Yan:2020, Sidenmark:2019, Sidenmark:2021}. 

Head-based gestures are another form of head-based interaction. They can be sensed and captured by the built-in inertial measurement unit sensor of smartphones or HMD. These gestures have been applied for different scenarios. Yi et al. \cite{Yi:2016} leveraged head gestures for user authentication. Rudi et al. \cite{Rudi:2016} used nod and shake to interact with maps on smart glasses. Tregillus \cite{Tregillus:2017} tracked head tilting (xz plane) for omnidirectional navigation in mobile VR. Yu et al. \cite{Yu:2019} explored the possibility of using move forward/backward (z axis) as a hands-free interaction approach in VR HMD, which has been applied for text entry recently \cite{Lu:2020}. Yan et al. \cite{Yan:2018} conducted an elicitation study and derived a head gesture set for HMD devices. They further evaluated its performance and concluded that head gestures could work as a supplementary input approach for HMD. 

In summary, prior work has proposed and examined using head gestures in HMD but outside of mode-switching. This study is inspired by this prior research and has adapted a proven method and task scenario to evaluate the use of head-based mode-switching gestures in VR HMD (Study One) and took a step further to assess the mode-switching gestures in a actual VR application with more realistic scenario (Study Two). To do this, we first sift out suitable head gestures based on specific design criteria. 

\section{Head-based Mode-Switching}
Head gestures are basic human movements and are quick, low-cost and fast to perform. There are a wide range of head gestures that can be translated into commands. Inspired by \cite{Surale:2019}, we defined the following three design criteria (DC) to narrow the scope for head-based mode-switching gestures in VR HMD.

\begin{itemize}
    \item \textbf{\textit{DC1}. User-maintainable}. A user-maintained mechanism helps users maintain awareness of the system state, in which potentially reduces mode errors \cite{Sellen:1992, Hinckley:2006}. Such awareness may also contribute to quickly correct a false positive if it happens. The head gestures should be able to be maintained by the users when the mode is enabled. This criterion rejects the gestures that are hard to perform or maintain. 
    \item \textbf{\textit{DC2}. Simple and fast}. Considering that users may be performing dual-hand interactions, their current task may already involve significant amount of cognitive and physical efforts. Thus, the mode-switching action need to be simple and straightforward so that it is possible to perform simultaneously with the existing tasks. In addition, it should be fast to allow a seamless integration with the other task(s). 
    \item \textbf{\textit{DC3}. Independent}. A mode-switching action should be independent of previous tracking states and should not rely on time-based actions \cite{Surale:2019}. This guarantees an explicit transition between states happens during mode-switching. Thus, gestures involving dwelling, combining different actions, or repeating an action are excluded. 
\end{itemize}

Based on above criteria, we derive 8 head gestures (in 4 pairs) that are potentially suitable for mode-switching in VR HMD, as shown in \autoref{fig:HeadGestures}. Pitch up/down, yaw left/right, and roll left/right are based on rotational movements along x, y, z axis, respectively. They are commonly performed in daily life and have been studied in prior work (e.g., \cite{Rudi:2016}). On the other hand, move forward/backward are based on positional movements along z axis. According to \cite{Yu:2019}, it is useful to allow users to lean slightly forward/backward to perform head motions along the z axis because it is more natural and helps with their balance. We also followed this approach in our studies. 

We conducted a pilot study with 6 participants to collect head movement data including the head position (Euler distance in 3 axes) and head orientation (Euler angles of pitch, yaw, roll). The participants performed each gesture five times. We then segmented each gesture and extracted the statistical characteristics of each gesture's potential position and orientation patterns in a certain duration. The information is used to setup each gesture's recognition. Our approach achieved near 100\% accuracy in gesture recognition using another 6 users in the pilot test.

\section{Study One: Feasibility Evaluation}
The aim of this user study was to investigate the following research question: \textit{which head gesture(s) can lead to better user performance and feedback in VR HMD mode-switching tasks?} To do this, we present a comparison of the above-mentioned eight head gestures in an extended line-drawing task in VR where they are used to switch between two colors when drawing the lines. 

\subsection{Participants and Apparatus}
Sixteen participants (8 females) from a local university volunteered to join this experiment. Participants were on average 21 years old (\textit{M} = 20.94, \textit{SD} = 1.88). According to the data from the pre-experiment questionnaire, all of them were right-handed, and had normal or corrected-to-normal vision and had no issues distinguishing colors. Six of them had used or interacted with VR HMD before the experiment. All participants had no physical discomfort and did not have problems moving their head to the mentioned positions and orientations. 

An HTC VIVE Pro VR HMD was used to provide the immersive virtual environments (VE). It has a resolution of 1440$\times$1600 px per eye, a refreshing rate of 90 Hz, and a 110$^\circ$ field of view. It was connected to an Intel Core i7 processor PC with NVIDIA Quadro P5200 graphics card. The VE was written in C\# using the Unity3D platform (version 2019.2.3f1). VIVE controllers were used to interact with the VE.

\subsection{Task}
In this study, we extended the VR line-drawing task described in \cite{Surale:2019} by asking participants to draw two lines with both hands at the same time. This task represented an extreme situation where users had to perform a non-trivial task with two hands simultaneously. The head gestures were used to change line colors (blue/red) which represented different modes. 

\begin{figure}[tb]
 \centering
 \includegraphics[width=\columnwidth]{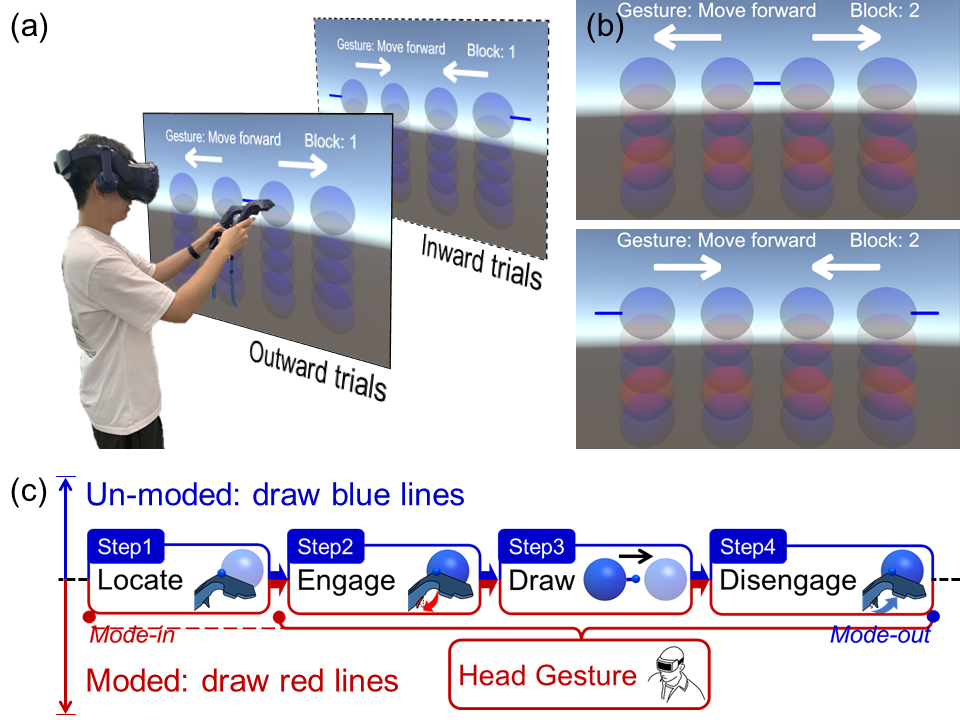}
 \caption{Experiment setup. (a) An example of a \textit{baseline} block which only requires to draw blue lines between the spheres. (b) \textit{Outward} (upper) and \textit{inward} trials (lower) in a \textit{compound} block. (c) The line-drawing process in a trial. The upper part (in blue) shows an \textit{un-moded} line-drawing process, and the lower part (in red) shows a \textit{moded} line-drawing process. The head gesture can be performed before engaging with the drawing activity, and should be maintained before disengaging from it.}
 \label{fig:S1_ExpSettings}
\end{figure}

At the beginning of the experiment, five groups of semi-transparent spheres were presented to participants in the VE (see \autoref{fig:S1_ExpSettings}a,b). Each group involved four spheres that had the same size (diameter = 80 mm) and was on the same horizontal line. The distance between two adjacent spheres was also 80 mm. The distance was chosen to allow participants see all the spheres and the text hints placed in front of them. Participants were required to draw two lines from the two inner spheres to the two outer spheres (\textit{outward trials}) or to do the reverse (\textit{inward trials}). To reduce any possible confusion, we put a line between the inner spheres for the outward trials and two lines outside the outer spheres for the inward trials. We also put two arrows above the spheres to show the direction of movement (see \autoref{fig:S1_ExpSettings}a,b). 

In each trial, the process of drawing could be divided into four steps (see \autoref{fig:S1_ExpSettings}c): (Step1) placing two controllers on the starting spheres; (Step2) engaging with the drawing task by pressing the trigger in each hand; (Step3) performing the drawing activity until the line reached the ending spheres; (Step4) releasing the triggers to disengage and indicate the end of the drawing task. If participants engaged or disengaged outside the specified spheres, an error was logged. Participants then needed to redo this trial. The spheres would turn opaque if the controllers were inside to provide visual feedback to users. 

We defined two types of task blocks: a \textit{baseline} block and a \textit{compound} block, which were based on the subtraction method \cite{Dillon:1990, Li:2005, Tu:2012, Surale:2017, Surale:2019}. Both types of blocks had two parts: the first part was to do 5 outward trials and the second part was 5 inward trials. In a baseline block, only a default mode was involved, and the spheres were in blue in all 10 trials (\autoref{fig:S1_ExpSettings}a). As such, participants were only required to draw \textit{un-moded}, blue lines. In a compound block, the spheres were in blue in the first, third, and last trial of each part, but in red in the second and fourth trials (\autoref{fig:S1_ExpSettings}b). For the former case, like in a baseline block, participants only needed to draw the un-moded lines (see the upper part in \autoref{fig:S1_ExpSettings}c). While for the latter case, participants had to switch the mode with the specified mode-switching head gesture and draw \textit{moded}, red lines (see the lower part in \autoref{fig:S1_ExpSettings}c). The head gesture should be performed before engaging with the drawing activity (before Step2). As suggested in previous studies \cite{Sellen:1992, Hinckley:2006, Surale:2019}, we used a user-maintained mechanism in this study. That is, users needed to hold the head gesture while drawing (Step2, 3, 4). There were text hints provided above the spheres to show the current head gesture and block (see \autoref{fig:S1_ExpSettings}a,b). If participants failed to switch in or out the mode, an error was logged, and they were required to redo this trial. All trials were performed in a standing position. 

\subsection{Design and Procedure}
The experiment included three sessions: (1) an introduction session to collect participants' demographics and background information and to brief them of study purpose, (2) an experimental session with eight sub-sessions for each head gesture, and (3) a post-experiment questionnaire session to collect their feedback. We used a within-subjects design with head gesture (\texttt{GESTURE}) as the independent variable. The order of \texttt{GESTURE} was counterbalanced using a Latin Square design. For each \texttt{GESTURE} condition, participants were requested to complete 9 blocks. The first block was a baseline task and then a compound task, alternating until the ninth block \cite{Li:2005}. The first two blocks were training blocks for participants to familiarize themselves with the head gesture and drawing process, and the remaining seven were formal blocks used in the analysis. Considering the duration of the experiments, we only asked participants to draw horizontal lines (outwards and inwards). When participants completed all the blocks for a \texttt{GESTURE} condition, they were given time to rest. A questionnaire about subjective ratings for the just-used head gesture was given during the break. After participants completed all the 8 conditions, we ask them to complete a post-questionnaire that asked them to provide a ranking of their preferred head gestures.

Based on this design, there were 8 \texttt{GESTURE} $\times$ 9 blocks of tasks $\times$ 10 line-drawing trials (5 outward + 5 inward) = 720 times of line-drawing for each participant, including 160 trials for training and 560 formal trials for analysis.

\subsection{Evaluation Metrics}
The dependent variables were \textit{mode-switching time} and \textit{error rates} in compound blocks. We used \texttt{BLOCK} to denote the compound blocks. \texttt{BLOCK} is a repeated measure in the analysis. The calculation of mode-switching time and the types of errors are explained in the following sub-sections. In addition, we used questionnaires to collect participants' subjective feedback, including ratings to each head gesture and a ranking of the most preferred gestures. 

\subsubsection{Mode-Switching Time}
\begin{figure}[tb]
 \centering
 \includegraphics[width=\columnwidth]{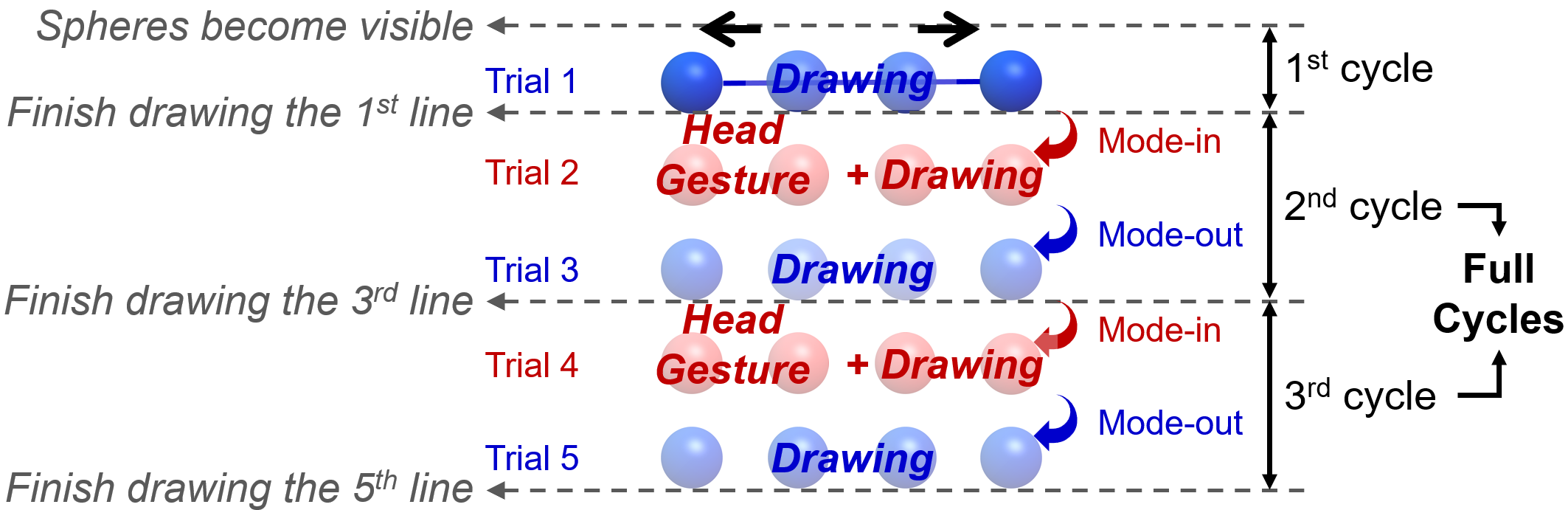}
 \caption{An example of the three cycles from outward trials in a compound block. A full cycle contains a complete mode-switching process: switching mode \textit{in} to draw \textit{moded} lines and switching mode \textit{out} to draw \textit{un-moded} lines.} 
 \label{fig:S1_Cycle}
\end{figure}

For each part of a block, there were three cycles involved, as shown in \autoref{fig:S1_Cycle}. The first cycle started from the moment the spheres were visible until the triggers were released after drawing the lines for the first group of spheres. The second cycle began immediately after, and ended when the triggers were released after drawing the lines for the third group. And, finally, the third cycle began immediately after, ending when the triggers were released after drawing the lines for the fifth group. 

The second and third cycles in each part of a block were defined as \textit{full cycles} \cite{Li:2005, Surale:2017, Surale:2019}. A full cycle in a compound block contained a complete mode-switching process including switching into a mode using the specified head gesture, completing a moded line-drawing task, switching back to the baseline mode, completing an un-moded line-drawing task. The first cycle guaranteed an un-moded line-drawing task performed before a full cycle. As such, there were 4 full cycles in each block (2 from outward trials, 2 from inward trials). 

We used the time of error-free trials for calculation. The mean time of error-free full cycles in a block was defined as \textit{average full cycle duration}. The mode-switching time for each of the 3 compound blocks was computed by subtracting the mean of the two adjacent baseline blocks’ average full cycle duration from the compound block’s average full cycle duration. In total, we derived 24 mode-switching time per participant (8 \texttt{GESTURE} $\times$ 3 \texttt{BLOCK}).

\subsubsection{Error Rates}
Like this prior study \cite{Surale:2019}, we defined four error types. First, a \textit{start error} happened if participants failed to engage in the line-drawing activity; that is, participants pressed the triggers outside the starting spheres. Second, an \textit{end error} occurred if participants failed to draw the lines to the specified end locations. In other words, participants released the triggers outside the ending spheres. Third, a \textit{mode-in error} happened if participants failed to use head gestures to draw the lines in red spheres, indicating a failure of switching into a mode. Finally, a \textit{mode-out error} happened if participants used a mode-switching gesture to draw the lines in the blue spheres, which would be a failure of switching back to baseline mode. Each of these error types were counted if they took place. In each \texttt{BLOCK}, an error rate was the ratio of the number of errors to the total number of trials, and an \textit{overall error rate} was the sum of all four error rates.

\subsubsection{Subjective Feedback}
After finishing each \texttt{GESTURE} condition, participants were asked to rate the just-used head gesture in terms of ease of learning, ease of use, accuracy, speed, neck fatigue and VR sickness. A 5-point continuous scale was used, with 1 being the most negative perception and 5 being the most positive. These questions were adapted from prior mode-switching studies \cite{Li:2005, Surale:2017, Surale:2019, Smith:2019, Pfeuffer:2020}. At the end of the experiment, we asked participants to rank their preferred head gestures based on the experience of using them for mode-switching in VR and provide their reasons.

\subsection{Results}
We used Excel (MS Office 365) and SPSS (version 26.0) for data processing and analysis. Before analyzing the data, we identified outliers of error-free line-drawing times in full cycles that were outside 3 standard deviations of the mean. We removed 1.56\% of the error-free trials in the full cycles. The remaining trials were used to calculate the mode-switching time with the explained strategy. 

Shapiro-Wilk tests showed that mode-switching time and overall error rates were normally distributed (\textit{p} $>$ .05). Thus, we performed Repeated Measures Analysis of Variance (RM-ANOVA) tests for these two measurements, Greenhouse-Geisser (\textit{$\epsilon$} $<$ .75 in our results) corrections when sphericity assumption was not met, and Bonferroni-adjusted post-hoc tests where applicable. Start, end, mode-in, and mode-out error rates were not normalised data according Shapiro-Wilk test (\textit{p} $<$ .05). As such, non-parametric Friedman tests were used for these measurements with \texttt{GESTURE} as the factor, and if a significant main effect was found, post-hoc analysis with Wilcoxon signed rank tests was conducted with Bonferrorni correction. 

\subsubsection{Mode-Switching Time}
\begin{figure}[tb]
    \centering
    \includegraphics[width = \columnwidth]{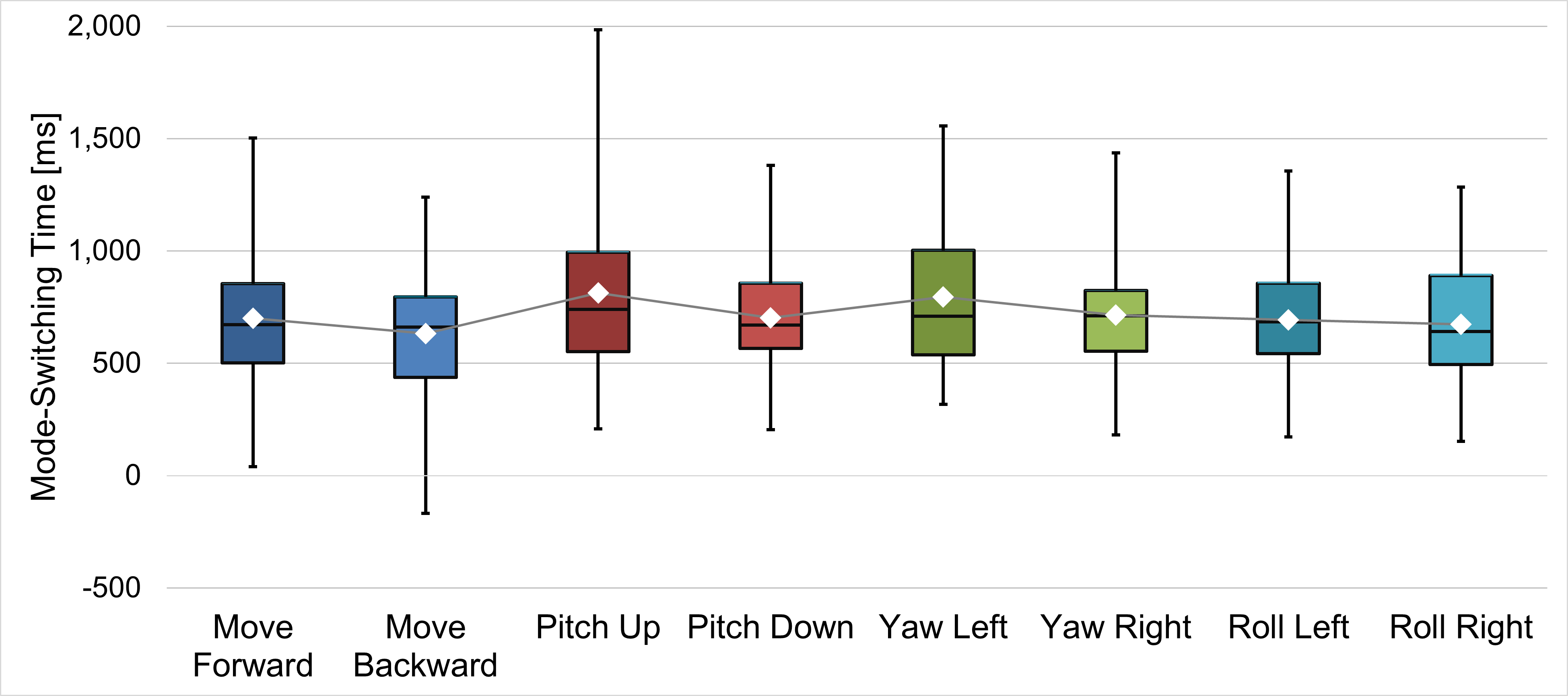}
    \caption{Boxplot of mode-switching time (in milliseconds) by \texttt{GESTURE}. White diamonds indicate mean mode-switching time for each head gesture.}
    \label{fig:S1_MSTBoxplot}
\end{figure}

RM-ANOVA test did not show a significant \texttt{GESTURE} $\times$ \texttt{BLOCK} interaction effect on mode-switching time (\textit{p} $>$ .05), indicating the stability of time performance among the blocks. This is in line with prior work \cite{Li:2005, Tu:2012, Surale:2017, Surale:2019}. \autoref{fig:S1_MSTBoxplot} shows the results of mode-switching time for each head gesture. There was no significant main effect for \texttt{GESTURE} on mode-switching time (\textit{p} $>$ .05). 

\subsubsection{Error Rates}
An error rates matrix summarizing the mean and standard deviation of error rates for each gesture is shown in \autoref{tab:errorrates}. RM-ANOVA test did not show a significant \texttt{GESTURE} $\times$ \texttt{BLOCK} interaction effect on overall error rates (\textit{p} $>$ .05), indicating the stability of accuracy among the blocks. There was a main effect for \texttt{GESTURE} on overall error rates (\textit{F}(7,105) = 2.281, \textit{p} = .033, \textit{$\eta^2$} = .132). However, the Bonferroni-corrected post-hoc tests did not yield significant differences (\textit{p} $>$ .05).

\begin{table*}[tb]
  \caption{Mean and standard deviation of overall, start, end, mode-in, and mode-out error rates (\%) by \texttt{GESTURE}. The lowest and second lowest error rates are highlighted in green and light green, while the highest and second highest error rates are highlighted in orange and light orange. }
  \label{tab:errorrates}
  \scriptsize%
	\centering%
  \begin{tabu}{%
	r%
	*{6}{c}%
	*{2}{r}%
	}
  \toprule
  Head Gesture & Overall Error Rate & Start Error Rate & End Error Rate & Mode-in Error Rate & Mode-out Error Rate \\
  \midrule
  Move Forward & 24.42$_{\pm15.35}$ & 2.52$_{\pm5.09}$ & \colorbox{YellowGreen}{14.70$_{\pm11.33}$} & \colorbox{YellowGreen}{1.74$_{\pm3.81}$} & \colorbox{Orange}{5.46$_{\pm9.74}$} \\
  Move Backward & \colorbox{Green}{17.71$_{\pm15.00}$} & 2.74$_{\pm5.49}$ & \colorbox{Green}{10.34$_{\pm11.68}$} & \colorbox{Green}{1.62$_{\pm3.36}$} & \colorbox{YellowOrange}{3.01$_{\pm5.27}$} \\
  Pitch Up & \colorbox{Orange}{28.98$_{\pm20.98}$} & \colorbox{YellowGreen}{2.45$_{\pm5.07}$} & 19.35$_{\pm14.80}$ & \colorbox{Orange}{5.94$_{\pm9.56}$} & 1.24$_{\pm3.80}$ \\
  Pitch Down & 28.36$_{\pm17.78}$ & \colorbox{Green}{1.29$_{\pm3.12}$} & 18.86$_{\pm13.67}$ & \colorbox{YellowOrange}{5.84$_{\pm8.43}$} & 2.37$_{\pm5.46}$ \\
  Yaw Left & 26.73$_{\pm21.42}$ & \colorbox{YellowOrange}{4.39$_{\pm8.65}$} & \colorbox{YellowOrange}{19.66$_{\pm18.04}$} & 2.20$_{\pm4.26}$ & \colorbox{Green}{0.49$_{\pm2.77}$}\\
  Yaw Right & \colorbox{YellowOrange}{28.43$_{\pm21.14}$} & \colorbox{Orange}{4.85$_{\pm8.69}$} & \colorbox{Orange}{20.68$_{\pm14.56}$} & 2.24$_{\pm4.49}$ & 0.66$_{\pm2.73}$\\
  Roll Left & \colorbox{YellowGreen}{22.42$_{\pm18.28}$} & 3.47$_{\pm5.39}$ & 15.65$_{\pm14.37}$ & 2.58$_{\pm6.52}$ & 0.71$_{\pm2.99}$\\
  Roll Right & 25.60$_{\pm14.31}$ & 3.27$_{\pm4.88}$ & 17.66$_{\pm12.04}$ & 4.14$_{\pm6.09}$ & \colorbox{YellowGreen}{0.53$_{\pm2.19}$}\\
  \midrule
  Total & 25.33$_{\pm18.42}$ & 3.12$_{\pm6.11}$ & 17.11$_{\pm14.2}$ & 3.29$_{\pm6.35}$ & 1.81$_{\pm5.17}$\\
  \bottomrule
  \end{tabu}%
\end{table*}

\begin{figure}[tb]
    \centering
    \includegraphics[width = \columnwidth]{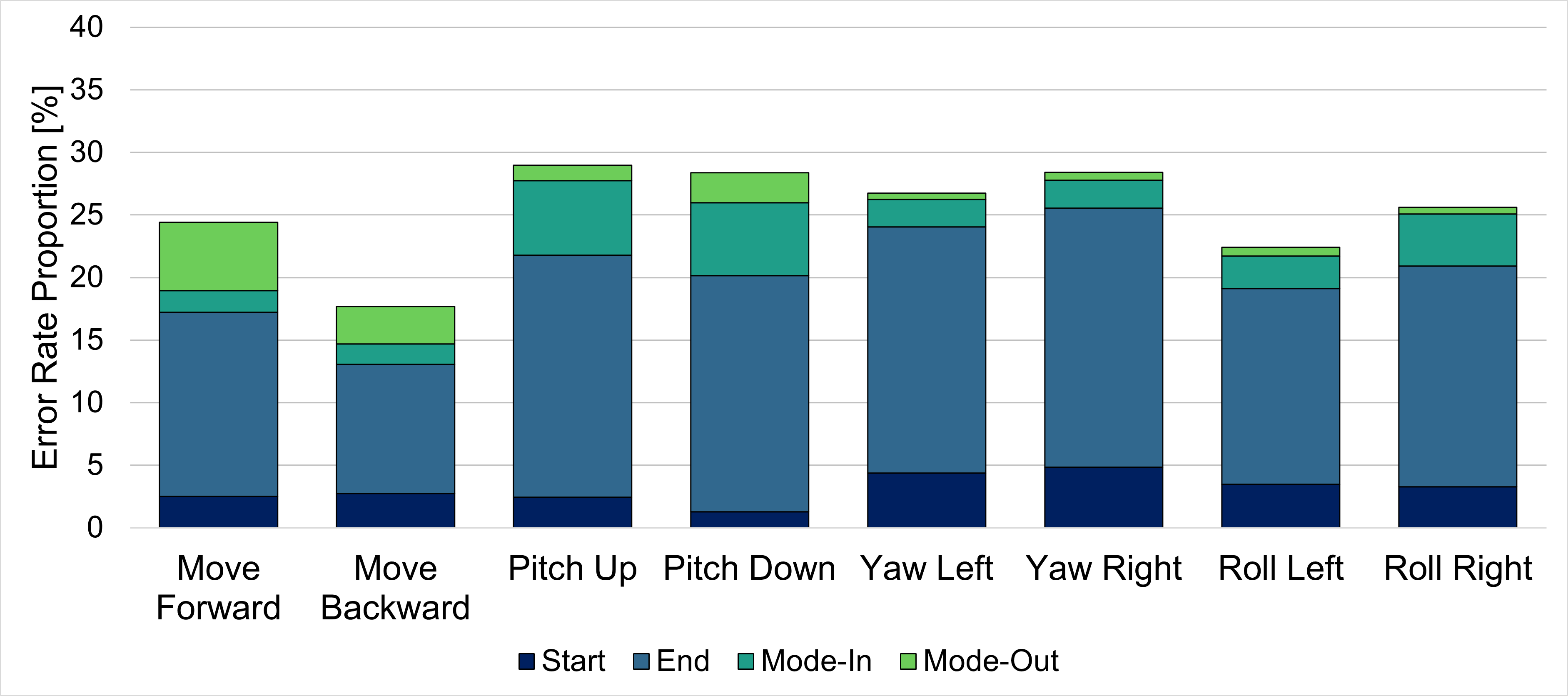}
    \caption{Proportion of each type of error rates by \texttt{GESTURE}. The sum of four types of error rates in a bar is the overall error rate for that head gesture.}
    \label{fig:S1_ErrorStack}
\end{figure}

\autoref{fig:S1_ErrorStack} illustrates the proportion of specific error rates by \texttt{GESTURE}. On average, end errors accounted for the largest weight (68\%). The proportion of start errors (12\%) and mode-in errors (13\%) were close, while mode-out errors had the smallest share (7\%). Results of the Friedman test showed no significant main effect for \texttt{GESTURE} on start error rates (\textit{p} $>$ .05). Average end error rates took up the largest proportion. The Friedman test yielded a significant main effect for \texttt{GESTURE} on end error rates (\textit{$\chi^2$}(7) = 31.318, \textit{p} $<$ .001). Post-hoc tests showed that move backward (10.34\%) had significant lower end error rates than pitch up (19.35\%), pitch down (18.86\%), yaw left (19.66\%), yaw right (20.68\%), and roll right (17.66\%) (\textit{p} $\leq$ .001). 

There was a significant main effect for \texttt{GESTURE} on mode-in error rates (\textit{$\chi^2$}(7) = 21.388, \textit{p} = .003). Post-hoc tests showed that move forward had significant lower mode-in error rates (1.74\%) than pitch down (5.84\%, \textit{p} = .006). Moreover, mode-in error rates of move backward (1.62\%) were significant lower compared to pitch up (5.94\%, \textit{p} = .002) and pitch down (5.84\%, \textit{p} = .001). There was a significant main effect for \texttt{GESTURE} on mode-out error rates (\textit{$\chi^2$}(7) = 40.215, \textit{p} $<$ .001). Post-hoc tests showed that yaw left (0.49\%) had lower rates than move forward (5.46\%, \textit{p} = .001), move backward (3.01\%, \textit{p} = .003), and pitch down (2.37\%, \textit{p} = .004). Besides, in addition to being higher than yaw left, move forward had a higher rates than yaw right (0.66\%, \textit{p} = .001), roll left (0.71\%, \textit{p} = .005), and roll right (0.53\%, \textit{p} $<$ .001). 

\subsubsection{Subjective Ratings, Ranking, and Feedback}
\begin{figure}[tb]
 \centering
 \includegraphics[width=\linewidth]{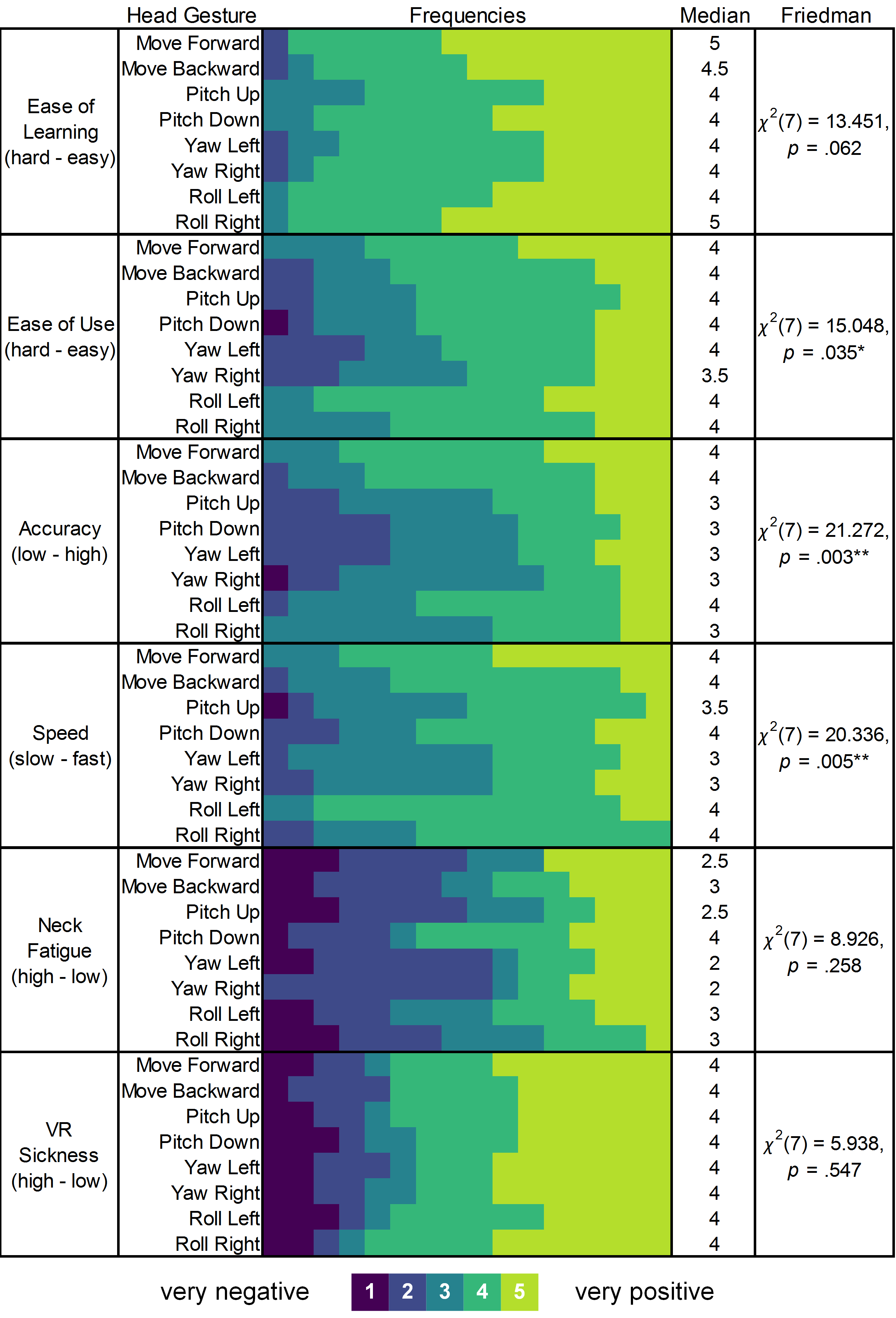}
 \caption{Subjective ratings and Friedman test results, indicating qualitative differences between the head gestures (* and ** represent significant results at .05, and .01, respectively).}
 \label{fig:S1_SubjectiveRatings}
\end{figure}

The results of the subjective ratings are listed in \autoref{fig:S1_SubjectiveRatings}. Friedman tests on subjective ratings showed significant main effects in ease of use (\textit{$\chi^2$}(7) = 15.048, \textit{p} = .035), accuracy (\textit{$\chi^2$}(7) = 21.272, \textit{p} = .003) and speed (\textit{$\chi^2$}(7) = 20.336, \textit{p} = .005), but no significant effects in the remaining measures (\textit{p} $>$ .05). In terms of ease of use, post-hoc tests did not yield any significant differences among the head gestures (\textit{p} $>$ .05). For accuracy, move forward was rated more accurate than pitch down (\textit{Z} = -2.877, \textit{p} = .004) and yaw right (\textit{Z} = -2.913, \textit{p} = .004). Moreover, move backward was rated more accurate than yaw right (\textit{Z} = -2.810, \textit{p} = .005). As for speed, post-hoc tests indicated that yaw left was rated faster than yaw right (\textit{Z} = -2.828, \textit{p} = .005).

\begin{figure}[tb]
 \centering
 \includegraphics[width=\columnwidth]{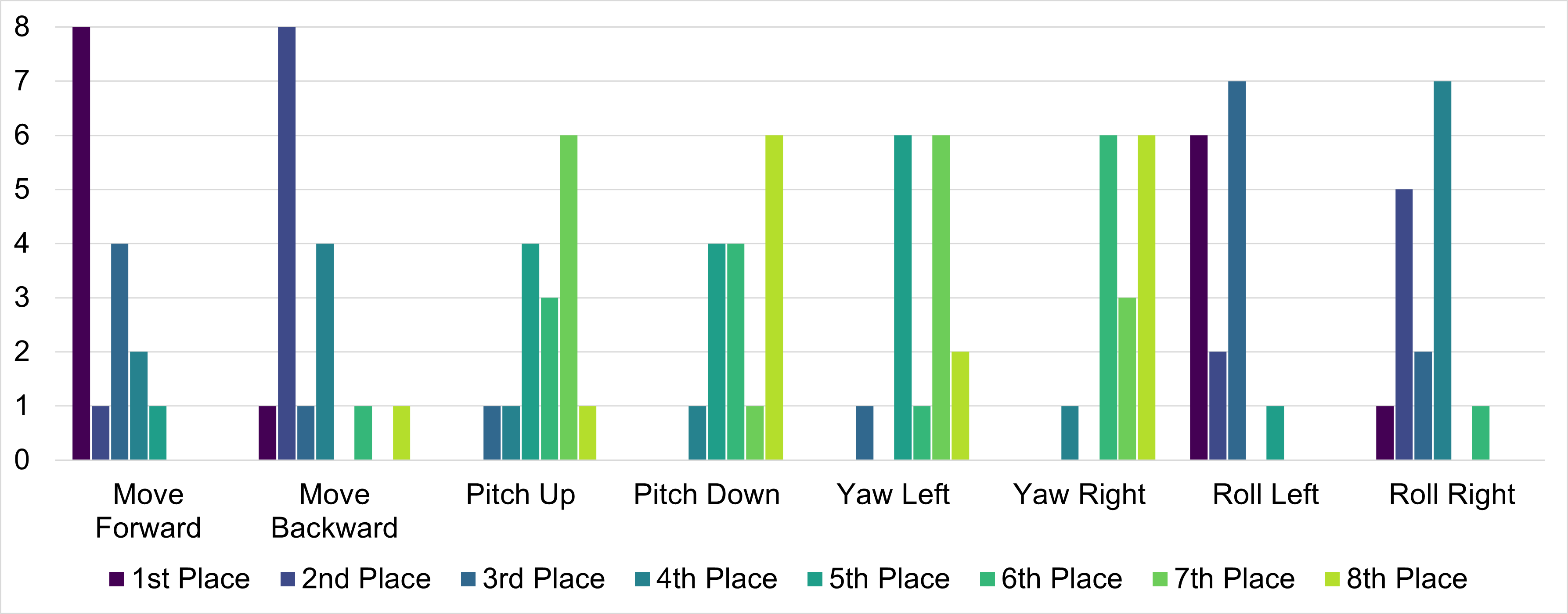}
 \caption{Distribution of participants' post rankings of the head gestures.}
 \label{fig:S1_Ranking}
\end{figure}

\autoref{fig:S1_Ranking} shows the distribution of participants' rankings of the head gestures after the trials. Participants tended to rank them based on gesture pairs. Paired gestures represented reverse pattern which could be a simple and intentional input \cite{Yan:2018} in case that more than one switching is required among multiple modes. Move forward/backward were ranked as the best head gestures for mode-switching in VR, closely followed by roll left/right. Pitch up/down and yaw left/right were ranked lower (see \autoref{fig:S1_Ranking}). Participants reported that, though they were able to complete the tasks, they could not see the whole task space when using pitch up/down or yaw left/right. 

\subsection{Discussion}
We investigated using potential head gestures to perform mode-switching in a dual-hand line-drawing task when using a VR HMD. Our study revealed insights into the user performance and feedback between these head gestures. 

Results did not yield a significant fast or slow head gesture for switching between modes in VR HMD. The time cost for performing these gestures was similar. All the eight filtered head gestures are onefold (\textbf{\textit{DC3}}) and simple to perform (\textbf{\textit{DC1}}). We used error-free trials for measuring mode-switching time. The average time was less than 850 ms for all motions, indicating that using head gestures for mode-switching in VR HMD is efficient if a gesture can be successfully performed. Our results meet \textbf{\textit{DC2}}.

We received high end error rates regardless of the head gestures (\textit{M} = 17.11\%) compared to prior VR mode-switching study \cite{Surale:2019}. In other words, participants faced difficulty in finishing the lines to the specified positions. One main reason might be the dual-hand line-drawing task we used was more challenging than the one-hand line-drawing task in VR \cite{Surale:2019} or pie-crossing tasks in 2DUI \cite{Li:2005,Tu:2012,Surale:2017}. Our setting represents an extreme situation where both hands are occupied. For such case, an explicit advantage of using head gestures for mode-switching is that its interaction is hands-free and does not rely on extra tracking devices. 

Move forward/backward were the two head gestures that participants gave the highest rankings and good ratings. Also, these two gestures had low overall error rates. The difference mainly comes from end errors where move backward and forward had the lowest and second lowest end error rates (see \autoref{tab:errorrates}). This was because the spheres were always in participants' front view, as they did not need them to do rotational movements when performing these two gestures. Move forward/backward had less mode-in errors but more mode-out errors than others, especially for move forward which had significantly higher mode-out errors. This implies that participants did not fully return to the default position, and thus did not switch back to the default mode. As mentioned, when performing these two head movements, the users' body may move slightly. Since our task occupied high cognitive resources, it was possible that participants were more focused on head movements, so they tried to move their head back but their body did not completely return. This can be easily minimized by fine-tuning the threshold to a slightly lower value for the returning motion. 

The objective results of roll left/right showed a sense of balance with little obvious positive and negative difference with other gestures in terms of mode-switching time and error rates. They received comparable high ranking and ratings with move forward/backward. Participants did not provide explicit reason for their ratings, especially in relation to move forward/backward. They said they were based on personal feelings of the gestures that were ``intuitive'' to make. In short, our results show an overall strong case for move forward/backward and roll left/right to be used for mode-switching in VR HMD. 

Pitch up/down, and yaw left/right received lower rank from participants (see \autoref{fig:S1_Ranking}) because they could not see the task area when performing these head gestures. Due to this situation, they perceived they had more mistakes with these gestures and gave lower ratings on accuracy compared to move forward/backward and roll left/right (see \autoref{fig:S1_SubjectiveRatings}). These results do not support the use of pitch up/down, and yaw left/right in the scenarios that have a continuous demand of users' visual attention on a certain area or direction of the VE. It is interesting to further evaluate these gestures in the case that the primary task would follow users' head movements; for example, performing selection in a menu which can be attached to users' view. Nevertheless, it is also possible to avoid these gestures because move forward/backward and roll left/right are still applicable to cases where the virtual objects move along with users' head motions/view.

In summary, our empirical comparison of the eight head gestures in a dual-hand line-drawing task point to four most suitable gestures to do mode-switching in VR HMD. They are move forward, move backward, roll left and roll right. These gestures have few error rates and better user's ratings and rankings compared to other head gestures. 

There are two aspects to note about Study One: (1) we only tested horizontal line-drawing to avoid long experiments which would lead to extra fatigue; and (2) while we did not asked participants not to move, they completed the tasks in a relatively static position and moved only for adjusting the distance between them and spheres. These two aspects led us to use a 3D-painting application to test these four gestures because it would involve omni- and multi-directional line-drawing and locomotion. Thus, with the initial insights from Study One, we conducted Study Two to further assess the four head gestures in a real VR painting application and investigate their applicability and integration in a natural and representative task.

\section{Study Two: Integration of the Head Gestures and Their Applicability in an Actual Application}
The second study aimed at evaluating how the four head gestures can be integrated in an actual application to allow users to switch modes while doing a somewhat more realistic task. We integrated the head gestures (move forward, move backward, roll left, and roll right) into a 3D-painting VR application to allow quick switches between the painting tools. We collected participants' subjective feedback to assess the gestures' performance, suitability, and usability. 

\subsection{Participants and Apparatus}
A total of 12 unpaid participants (5 females) were recruited to do this experiment. They were aged between 19 and 29 years (\textit{M} = 22.92, \textit{SD} = 3.15). All of them were right-handed. Five of them draw or paint frequently (i.e., at least once a month). Nine participants were regular VR users and four reported prior experience of drawing in VR. We used the same apparatus as in the previous study.

\subsection{Application Scenario}
\begin{figure}[tb]
 \centering
 \includegraphics[width=\columnwidth]{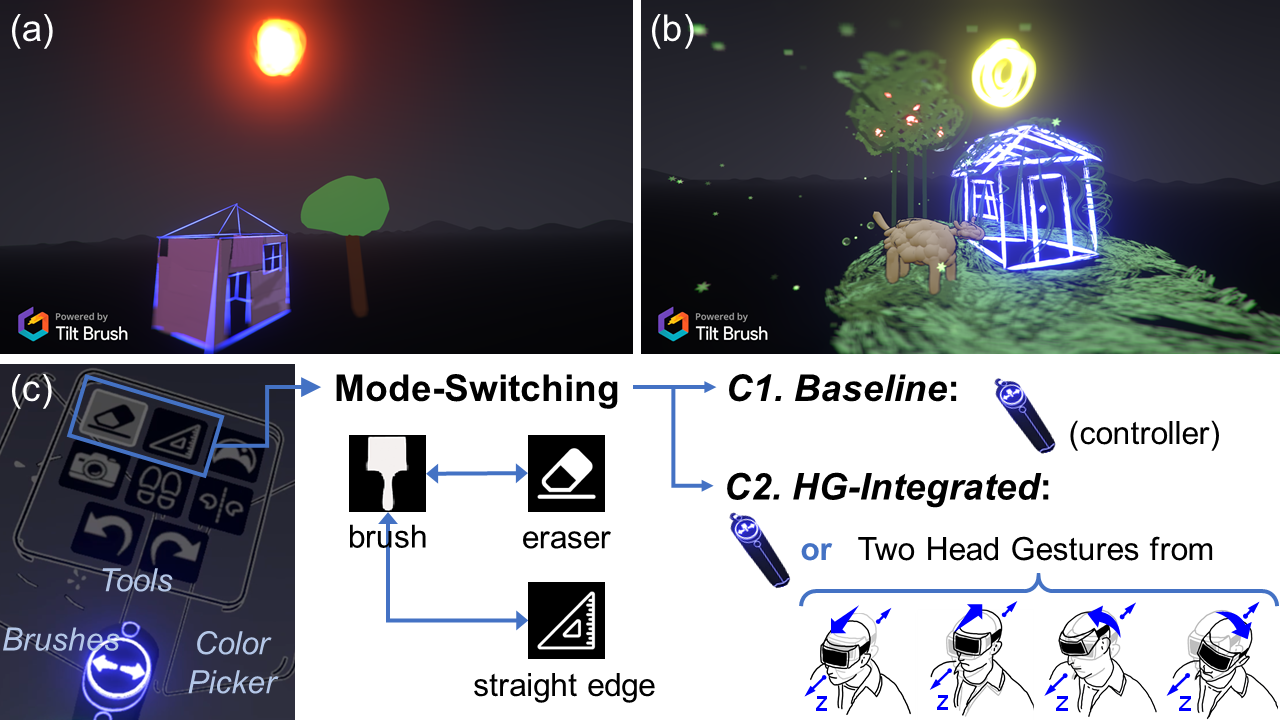}
 \caption{(a) and (b) Two art work from participants (P6, P9). (c) Experimental conditions. \textit{C1.Baseline}: use controller to switch. \textit{C2.HG-Integrated}: allow quick switch using head gestures. Note that eraser and straight edge are both located in \textit{Tools} interface. }
 \label{fig:Study2}
\end{figure}

We used a 3D-painting scenario to further evaluate the four head-based mode-switching gestures. An open-source application, Tilt Brush\footnote{https://www.tiltbrush.com/} was used to provide the virtual painting environment. Participants were requested to draw a courtyard in 3D space. We suggested that participants draw a sun, a house, and a tree but the actual appearance of these objects can be freeform (see \autoref{fig:Study2}a,b). The controller held by dominant hand was for drawing and non-dominant one was a tool panel, which had three-cornered interfaces including \textit{Tools}, \textit{Color Picker}, and \textit{Brushes}, as shown in \autoref{fig:Study2}c. During the drawing process, participants were free to use the built-in tools on these interfaces. We integrated the four head gestures into the application. They could be used to achieve quick switch between drawing and erasing, and between drawing free lines and drawing straight lines; that is, switch to use an eraser (and back to drawing) or to draw with the help of a straight edge (and back to free drawing). Both eraser and straight edge are placed on \textit{Tools} interface by default. We offered participants an option to choose two preferred head gestures for the just-mentioned two types of switching after experiencing and comparing them in the VE. The integrated head gestures did not replace the default operation. In other words, participants could still switch the mode by selecting the options from the tool panel, if they wanted to (see \autoref{fig:Study2}c).

\subsection{Design and Procedure}
We first collected participants' demographic and background information via a questionnaire. Then, participants were briefed about the research aim and introduced to painting in VR using Tilt Brush with a short tutorial about the usage of its common tools. They got at most 10 minutes to familiarize themselves with the operations in the VE. After this, we introduced the four head gestures to them. Participants could try the head gestures and use them to switch between tools in the application. They needed to select two of gestures that they thought practical and suitable for switching between modes. Then the formal experiment started. It had two sessions, one using the non-modified version of the application (\textit{Baseline}), and another using the version integrated with head gestures (\textit{HG-Integrated}). Their order was alternated. Each session lasted about 10 minutes, where participants could make their own creation that were related to the topic of a courtyard. After completing each session, we gave them a short version of User Experience Questionnaire (UEQ) \cite{UEQ}. At the end of the experiment, we conducted a structured interview asking their experience with the two versions. 

\subsection{Results and Discussion}
\subsubsection{UEQ results}
\autoref{tab:UEQ} summarizes the results from UEQ in terms of pragmatic, hedonic, and overall quality. The average scores in \textit{HG-Integrated} are all higher than in \textit{Baseline}. Results from Wilcoxon signed rank tests showed that the differences in UEQ scores between \textit{HG-Integrated} and \textit{Baseline} were significant in terms of \textit{overall} quality (\textit{Z} = -2.394, \textit{p} = .017) and \textit{hedonic} quality (\textit{Z} = -2.937, \textit{p} = .003), but not significant in \textit{pragmatic} quality (\textit{p} $>$ .05). The results indicate that using head gestures to support mode-switching in the VR drawing application provides better user experience.

\begin{table}[tb]
  \caption{Mean and standard deviation of the results from UEQ for pragmatic, hedonic, and overall quality. Higher scores indicate better user experience.}
  \label{tab:UEQ}
  \scriptsize%
	\centering%
  \begin{tabu}{%
	r%
	*{4}{c}%
	*{1}{r}%
	}
  \toprule
  Condition & Pragmatic & Hedonic & Overall \\
  \midrule
  \textit{Baseline} & 0.29$_{\pm1.25}$ & -0.29$_{\pm1.19}$ & 0$_{\pm1.13}$\\
  \textit{HG-Integrated} & 1.06$_{\pm1.00}$ & 1.42$_{\pm0.71}$ & 1.24$_{\pm0.72}$\\
  \bottomrule
  \end{tabu}%
\end{table}

\subsubsection{Selection of Head Gestures}
In this study, we allowed participants to select their two preferred head gestures and match them with the two mode-switching process after experiencing them in the application. Such gesture-to-action mapping represents a sense of customization that is commonly seen in software applications. 

Most participants (\textit{N} = 9) selected paired gestures. Seven of them chose roll left/right for eraser or straight edge. Four commented that move forward/backward were difficult or uncomfortable to perform while drawing. Two participants (P3, P6) worried that they might lose their balance while performing or maintaining move forward/backward. On the other hand, two participants (P4, P7) chose move forward/backward. P4 mentioned that: \textit{``move forward/backward were more intuitive and easier to perform.''} While P7 felt hard to maintain roll left/right. 

Three participants (P1, P8, P10) selected mixed settings. They all chose roll left for straight edge and move forward for eraser. They felt roll left/right might affect drawing due to the rotation of the vision, and thus did not use them for straight edge. In addition, they thought roll left was more natural and intuitive compared to roll right. They commented that move forward was ``\textit{more convenient}'' and ``\textit{easier}'' than move backward. In general, these comments were in line with the subjective feedback from Study One---users have different perceptions on the head gestures, leading to unique preferences. 

\subsubsection{Participants' Comments}
All 12 participants commented that they felt mode-switching in \textit{HG-Integrated} was more efficient than \textit{Baseline}. They used ``\textit{continuous}'' and ``\textit{smooth}'' to describe the drawing with the former, and ``\textit{cumbersome}'' and ``\textit{interruptive}'' for the baseline. The head-based gestures improved the mode-switching procedure by saving the time to locate the target mode and default mode before and after using it. Participants (P6, P9, P10) noted that head-based mode-switching was particularly efficient if the current interface was Brushes or Color Picker (see \autoref{fig:Study2}), in which by default an additional switch between the upper-level interfaces was required. P2 and P4 mentioned that direct switching using head gestures helped to get more chances for resting their non-dominant hands, reducing arm fatigue. 

Almost all participants (\textit{N} = 11) agreed that learning the head-based mode-switching approach in the drawing scenario was effortless. This extends the results we obtained from Study One showing that the head gestures for mode-switching were easy to learn (see \autoref{fig:S1_SubjectiveRatings}). P5 said that the customization of the head gestures before the formal experiment improved the learnability and memorability. 

Some participants (\textit{N} = 6) believed the experience and the efficiency of using head gestures to do mode switching in the application could be further improved if the gestures can be more resistant to false positives, though they did not happen frequently as we only observed this in a handful of cases. To complete the 3D-painting task, participants inevitably needed to walk around, which we allowed them to do, leading to more locomotion than standing without much motion like in Study One, though we did not specifically asked them not to move. This shows that head gestures can work reliably even when participants are in motion. Also, three participants (P4, P7, P12) suggested having visual cues so that they could know the switching of modes. Providing visual cues, voice prompts, or tactile feedback can help users be aware of the current mode.  

Two participants (P3, P10) thought their judgement or estimation of objects in some application scenarios might be influenced while performing roll left/right. They also thought that with some more practice, this should not be a major issue. Similarly, P5 and P11 commented that head gestures may be more difficult to perform when they need to stand on their toes or squat. They said that these activities may not be frequent in VR applications and, at the end, would not affect the main usefulness of head gestures to do mode switching.

\section{Summary and Design Recommendations}
In this section, we summarize the results from the two user studies, and distill design recommendations for incorporating head-based gestures in VR HMD.

\textbf{Head gestures can support mode switching in VR}. We derived four head gestures that can be suitable for mode-switching in VR from the first controlled experiment. To understand their suitability and usability for actual applications, we conducted a second user study where these gestures were embedded in an existing application. Based on the results from two studies, incorporating head gestures for mode-switching in VR is an efficient hands-free approach to provide a ``continuous'' and ``smooth'' experience. Given that head gestures can be detected effectively and performed easily \cite{Yan:2018,Yu:2019}, they complement other modalities and tasks. 

\textbf{Head gestures should be customized by end-users}. We noticed that participants had different preferences on head gestures from the first study (see \autoref{fig:S1_Ranking}). As such, we gave participants in the second study opportunities to try and select preferred gestures based on their experience before the actual use. We received positive feedback about customizing their own gestures. Therefore, allowing users to select their preferred gestures would be helpful for improving their learnability and usability.

\textbf{Selecting suitable head gestures according to the use case}. In the first study, we found that move forward/backward had better performance than roll left/right and were more preferred by participants for mode-switching in VR. Despite this, a small number of participants in the second study selected to help them make mode switches. This choice could be due to the different nature of the tasks in the experiments. The more controlled setting in the first study represent a single task, while the painting task in the second study involved multitasking which was more dynamic and complex. In this sense, designers should consider how the unique features of the head gestures might affect user experience in a specified application.

\section{Limitations and Future Work}
This research has the following limitations, which could represent avenues for future work. First, we used a simple strategy for sensing and capturing head gestures. Though it had high accuracy, it can be further improved by optimizing the algorithm and testing different sensing parameters. This was not the concern of the current study but it could be relevant for a wider adoption of head gestures in other types of HMD. Second, we only focused on a user-maintained mechanism for mode-switching. Though a system-maintained mechanism may be error prone \cite{Sellen:1992,Hinckley:2006}, it could still be worthy to examine its comparative performance and usability. Furthermore, it could be also interesting to design and evaluate a hybrid method for real applications. Third, we mainly investigated head-based switching between two modes. While some scenarios could involve more than two modes, the two paired gestures could cover a wider range of practical scenarios in VR applications. In the future, we plan to explore the feasibility and applicability of having more than two modes for each pair of head gestures. One approach is to have multiple modes placed in a continuum. By using gestures like move forward/backward, one can traverse through the available modes and returning to a neutral position represents the selection of the mode. Forth, we focused on exploring and evaluating different head gestures for hands-free mode switching. We plan to conduct user studies to further compare the usability of head-based mode switching with other hands-free approaches (like gaze) in the future. 

\section{Conclusion}
In this paper, we explored the use of head gestures for mode switching in Virtual Reality head-mounted displays (VR HMD). We first derived head gestures that are potentially suitable based on three design criteria. We then conducted a user study to systematically compare their efficiency, accuracy, and usability. In a second study, we further investigated the usability of the four most usable head gestures in an actual 3D-painting application. Results from two studies show that move forward, move backward, roll left and roll right offer relatively better performance and received the best user ratings. They also show that they can be easily integrated into an application and learned by users. Based on our results, we conclude that head gestures can support mode switching in VR and represent a suitable approach that is hands-free, fast, accurate, and low-cost in current VR HMD. Our work contributes a first systematic exploration of using head gestures for mode switching in VR HMD, a set of suitable gestures, and several design recommendations for their use in VR applications.

\acknowledgments{
The authors wish to thank the participants for their time and the reviewers for their insightful comments that have helped improve the paper. This work was supported in part by Jiaotong-Liverpool University (XJTLU) Key Special Fund (No: KSF-A-03) and the National Research Foundation, Singapore under its AI Singapore Programme (AISG Award No: AISG2-RP-2020-016).}

\bibliographystyle{abbrv-doi}

\bibliography{template}
\end{document}